\documentclass[12pt]{article}
\usepackage{amsmath,amssymb,amscd,cite}
\newtheorem{thm}{Theorem}[section]

\newtheorem{lem}[thm]{Lemma}
\newtheorem{cor}[thm]{Corollary}

\newtheorem{defi}[thm]{Definition}
\newcommand{\qed}{\hfill $\Box$ \\}
\font\msbm=msbm10 at 12pt
\newcommand{\Z}{\mbox{\msbm Z}}

\newcommand{\F}{\mbox{\msbm F}}

\newtheorem{ex}[thm]{Example}

\date{}

\begin{document}

\title{The Dual and the Gray Image of Codes over $\mathbb{F}_{q}+v\mathbb{F}_{q}+v^{2}\mathbb{F}_{q}$}
\author{A. Melakheso and K. Guenda}
\maketitle
\begin{abstract}
In this paper, we study the linear codes over the commutative ring $R=\F_{q}+v\F_{q}+v^{2}\F_{q}$ and their Gray images, where $v^{3}=v$. We define the Lee weight of the elements of $R$, we give a Gray map from $R^{n}$ to $\F^{3n}_{q}$ and we give the relation between the dual and the Gray image of a code. This allows us to  investigate the structure and properties of self-dual cyclic, formally self-dual and the Gray image of formally self-dual codes over  $R$. Further, we give several constructions of formally self-dual codes over $R$.

\end{abstract}
\section{Introducton}
Codes over finite rings have been studied since the early. There are a lot of work on this codes after the discovery of certain good non-linear codes can be constructed from cyclic codes over $\Z_{4}$. Recently, Zhu et al. considered linear codes over finite non-chain ring $\F_{q}+v\F_{q}$, they investigated a class of constacyclic over $\F_{p}+v\F_{p}$\cite{Zhu}. \\
In \cite{Jia} the authors studied cyclic codes and the weight enumerator of linear codes over $\F_{2}+v\F_{2}+v^{2}\F_{2}$. 

In this paper, we focus on codes over the ring  $R=\F_{q}+v\F_{q}+v^{2}\F_{q}$, where $v^{3}=v$. The remainder of the paper is organized as follows. In section 2, we give some basic knowledge about the finite ring $R=\F_{q}+v\F_{q}+v^{2}\F_{q}$. In section 3, we define the Lee weight of the element of $R$, and introduce a Gray map. This map leads to some useful results on linear codes over $R$. We investigated the relation between the dual and the Gray image of codes. In section we study cyclic codes over $R$. We finish by giving different constructions of formally self-dual and we study the Gray map of formally self-dual codes over $R$.
\section{Preliminaries}
In this section, we introduce some basic results on linear codes over the ring $R=\F_{q}+v\F_{q}+v^{2}\F_{q}$, $q$ a prime power, where $v^{3}=v$. An element $x$ of $R$ can be expressed uniquely as $x=a_{0}+va_{1}+v^{2}a_{2}$, where $a_{i}\in \F_{q}, i=0, 1, 2$.
$R=\F_{q}+v\F_{q}+v^{2}\F_{q}$ is principal ideal ring and has three non-trivial ideals, namely,
\begin{center}
$<v>=\lbrace a_{1}v;a_{1}\in \F_{q}\rbrace$,

$<1-v>=\lbrace a_{2}(1-v);a_{2}\in \F_{q}\rbrace$,

$<1-v^{2}>=\lbrace a_{3}(1-v^{2}); a_{3}\in \F_{q}\rbrace$.
\end{center}
Let $(\F_{q}+v\F_{q}+v^{2}\F_{q})^{n}$ be the $\F_{q}+v\F_{q}+v^{2}\F_{q}$-module of $n$-tuple over $\F_{q}+v\F_{q}+v^{2}\F_{q}$. A linear code $C$ of length $n$ over $R$ is an $R$-submodule of $R^{n}$. An element of $C$ is called a codeword of $C$. A generator matrix of $C$ is a matrix whose rows generate $C$. The Hamming weight $w_{H}(c)$ of codeword $c$ is the number of nonzero components in $c$. The Hamming distance $d_{H}(C)$ of $C$ is defined as $d_{H}=min\lbrace w_{H}(x_{1}-x_{2})\left\vert x_{1}, x_{2}\in C, x_{1}\neq x_{2}\rbrace\right.  $. Let  $x=(x_{0}, x_{1},...,x_{n-1})$ and $y=(y_{0}, y_{1},...,y_{n-1})$ be two element of $R^{n}$, the Euclidean inner product is given as:
\begin{center}
$x.y=x_{0}y_{0}+x_{1}y_{1}+...+x_{n-1}y_{n-1}$.
\end{center}
The dual code $C^{\perp}$ of $C$ with respect to the Euclidean inner product is defined as
\begin{center}
$C^{\perp}=\lbrace x\in R^{n} \left\vert  x.y=0, \forall y\in R^{n}\rbrace \right. $.
\end{center}
$C$ is self-dual if $C=C^{\perp}$, $C$ is self-orthogonal if $C\subseteq C^\perp$\cite{Liu}.
 Let $ C $ be linear code of length $ n $ over $ R$.
 Define
 \begin{center}
 $ \begin{array}{cccc}
C_{1} &=&\left\{ a\in \F_{q}^{n}; \exists b,c\in \F_{q}^{n}; a+vb+v^{2}c\in C\right\} ,\\
C_{2} &=&\left\{ a+b\in \F_{q}^{n};\exists c\in \F_{q}^{n};a+vb+v^{2}c\in C\right\}, \\
C_{3} &=&\left\{ a+b+c\in \F_{q}^{n};a+vb+v^{2}c\in C\right\}.
\end{array}$
 \end{center}

 Obviously, $ C_{1} $, $ C_{2} $, and $ C_{3} $ are linear codes over $\F_{q}$.\\
  By the definition of $ C_{1} $, $ C_{2} $, and $C_{3}$ we have that $C=vC_{1}\oplus \left( 1-v\right) C_{2}\oplus \left( 1-v^{2}\right) C_{3}$. So $C_{1},$ $C_{2}$ and $C_{3}$ are unique. That we have $\left\vert C\right\vert =\left\vert C_{1}\right\vert \left\vert C_{2}\right\vert\left\vert C_{3}\right\vert$.
\section{Gray Map over $R$}
 Let $ x=a_{0}+va_{1}+v^{2}a_{2} $ be an element of $R$ where  $a_{i}\in \F_{q}, i=0, 1, 2  $.\\
 We define Gray map $\Psi$ from $R$ to $\F^{3n}_{q} $ by

\begin{center}
 $\Psi :R  \rightarrow \F^{3n}_{q} $\\
 $ x=a_{0}+va_{1}+v^{2}a_{2} \mapsto  \Psi\left( x\right) =\left(a_{0}, a_{0}+ a_{2}, a_{1} \right) $
\end{center}
From definition, the Lee weight of elements of $ R $ is defined as follows\\
\begin{equation*}
w_{L}\left(a_{0}+va_{1}+v^{2}a_{2}\right) =\left\lbrace
\begin{array}{ccccc}
0  & if  & a_{0}=0;&a_{1}=0;&a_{2}=0\\
1  & if  & a_{0}=0;&a_{1}\neq 0;&a_{2}=0\\
1   & if  & a_{0}\neq 0;&a_{1}\neq 0; &a_{2}=0\\
1 & if & a_{0}\neq 0;&a_{1}=0; &a_{0}+a_{2}=0\left[ mod \right] \\
1 & if & a_{0}=0;&a_{1}\neq 0;&a_{2}\neq0\\
2  & if  & a_{0}\neq0;&a_{1}= 0;&a_{0}+a_{2}=0\left[ mod \right]\\
2 & if& a_{0}=0;&a_{1}\neq 0;&a_{2}\neq 0\\
2 & if& a_{0}\neq 0;&a_{1}\neq 0;&a_{0}+a_{1}=0\left[ mod \right]\\
3 & if & a_{0}\neq 0;&a_{1}\neq 0;&a_{0}+a_{2}=0\left[ mod \right]\\
3 & if  & a_{0}\neq 0;&a_{1}\neq 0;&a_{2}=0\\

\end{array}
\right. \end{equation*}
For any codeword $c=\left( c_{0}, c_{1},..., c_{n-1}\right) $ the Lee weight of $ c $ is defined as $w_{L}\left( c\right) =\overset{n-1}{\underset{i=0}{\sum w_{L}\left( c_{i}\right) }}$ and the Lee distance of $ c $ is defined as $ d_{L}\left( c\right) =min d_{L}\left( c,\acute{c}\right)$, where $d_{L}\left( c,\acute{c}\right)=w_{L}\left( c-\acute{c}\right)$ for any $\acute{c}\in C$, $ c\neq \acute{c}$.\\
Now can be extended the Gray map to $R^{n}$.
\begin{defi}
The Gray map $\Psi$ from $ R^{n}$ to  $ \F^{3n}_{q} $ is define $\Psi\left(a_{0}, a_{1}, a_{2}\right) =\left( a_{0}, a_{0}+a_{2}, a_{1}\right) $ for all $ a_{i}\in \F^{3n}_{q}$, $ i=0, 1, 2 $.
\end{defi}
\begin{thm}
 The Gray map $ \Psi $ is a  weight preserving map from $ \left( R^{n}, \textit{Lee weight} \right)$ to\\
  $ \left(\mathbb{F}^{3n}_{q}, \textit{Hamming weight} \right)$.
\end{thm}
\begin{pf}

 Let $c=\left( c_{0}, c_{1},..., c_{n-1}\right)\in R^{n} $  and $\acute{c}=\left( \acute{c}_{0}, \acute{c}_{1},..., \acute{c}_{n-1}\right) \in R^{n}$. Since $w_{L}\left(c \right) = w_{H}\left(\Psi\left(c_{i} \right)  \right)$, $ i=0, 1,..., n-1 $, then we have
\begin{center}
$w_{L}\left( c\right) =\sum \limits_{i=0}^{n-1}w_{L}\left( c_{i}\right) =\sum \limits_{i=0}^{n-1} w_{H}\left( \Psi\left( c_{i}\right) \right)=w_{H}\left( \Psi\left( c\right) \right)$.

\end{center}
\qed
\end{pf}

\begin{thm}
If $ C $ is self orthogonal, so is $ \Psi\left( C\right)$.
\end{thm}
\begin{pf}
Let $ c=a_{1}+v b_{1}+v^{2}c_{1}$, $ \acute{c}=a_{2}+v b_{2}+v^{2}c_{2}$, where $ a_{1}, b_{1},c_{1}, a_{1},c_{1} ,b_{1}, a_{1}, b_{1},c_{1} \in \F_{q}$. And let
\begin{center}
$c\cdot\acute{c}=a_{1}a_{2}+v\left( a_{1}b_{1}+b_{1}a_{2}+b_{1}c_{2}+c_{1}b_{2}\right) +v^{2}\left( a_{1}a_{2}+b_{1}b_{2}+c_{1}a_{2}+c_{1}c_{2}\right)$
\end{center}

if $ C $ is self orthogonal, so we have $a_{1}a _{2}=0$, $a_{1}b_{1}+b_{1}a_{2}+b_{1}c_{2}+c_{1}b_{2}=0$ and $ a_{1}a_{2}+b_{1}b_{2}+c_{1}a_{2}+c_{1}c_{2}=0$.\\
 From
\begin{center}
$ \Psi\left( c\right) \cdot\Psi\left( \acute{c}\right)=\left( a_{1}, a_{1}+c_{1}, b_{1}\right)\left( a_{2}, a_{2}+c_{2}, b_{2}\right) $ $ = a_{1}a_{2}+a_{1}a_{2}+a_{1}c_{2}+c_{1}a_{2}+c_{1}c_{2}+b_{1}b_{2}=0$.
\end{center}
Therefore, we have $ \Psi\left( C\right)  $ is self orthogonal.
\qed
\end{pf}
\begin{cor}
If $ C $ is a linear code over $ R $, the minimum Lee weight of $C$ is the same as the minimum Hamming weight of $ \Psi\left( c\right) $.
\end{cor}
Let $ d_{L} $ minimum Lee weight of linear code over $R$. \\
Then $ d_{L} =min\left\lbrace d_{H} \left( C_{1}\right),  d_{H} \left( C_{2}\right), d_{H} \left( C_{3}\right)\right\rbrace  $ where $ d_{H} \left( C_{i}\right) $ denotes the minimum Hamming weight of codes, $ C_{1} $, $ C_{2} $ and $ C_{3}$.
\begin{lem}\label{lem:1}
Let $C=vC_{1}\oplus\left( 1-v\right) C_{2}\oplus \left( 1-v^{2}\right) C_{3}$ be a linear code over $R$, where $ C_{i} $ is a linear code with dimension $ k_{i} $ and minimum Hamming distance $d\left( C_{i}\right) $ for $ i=1,2,3 $.
 Then $ \Psi\left( C\right)  $ is a linear code with parameters
$ \left[ 3n, k_{1}+k_{2}+k_{3},min\left\lbrace d \left( C_{1}\right), d \left( C_{2}\right) ,d\left( C_{3} \right) \right\rbrace \right]  $ over $ \F_{q} $.
\end{lem}
In the following we investigate the relation between the dual and the Gray image of a code over $R$.
\begin{thm}
If $ C^{\perp} $ is the dual of $ C $, then $ \Psi\left(C \right) ^{\perp}=\Psi\left(C ^{\perp}\right) $. Moreover, if $ C $ is self-dual code, so is $ \Psi\left(C \right) $.
\end{thm}
\begin{pf}

For all $c_{1}=a_{0}+va_{1}+v^{2}a_{2}\in C$ and $c_{2}=\acute{a}_{0}+\acute{a}
_{1}v+\acute{a}_{2}v^{2}\in C^{\bot },$ where $a_{i},\acute{a}_{i}\in \F_{q}^{n},i=0, 1, 2$, if $c_{1}\cdot c_{2}=0$, then we have

$ c_{1}\cdot c_{2}=a_{0}\acute{a}_{0}+v\left( a_{0}\acute{a}_{1}+a_{1}\acute{a}_{0}+a_{1}\acute{a}_{2}+a_{2}\acute{a}_{1}\right) +v^{2}\left( a_{0}\acute{a}_{2}+a_{1}\acute{a}_{1}+a_{2}\acute{a}_{0}+a_{2}\acute{a}_{2}\right) =0$.
Implying that $a_{0}\acute{a}_{0}=0, a_{0}\acute{a}_{1}+a_{1}\acute{a}_{0}+a_{1}\acute{a}_{2}+a_{2}\acute{a}_{1}=0$, and $a_{0}\acute{a}_{2}+a_{1}\acute{a}_{1}+a_{2}\acute{a}_{0}+a_{2}\acute{a}_{2}=0$.
Therefore
\begin{center}
$ \Psi \left( c_{1}\right) \cdot \Psi \left( c_{2}\right) =\left(a_{0},a_{0}+a_{2},a_{1}\right)\cdot\left( a_{0},\acute{a}_{0}+\acute{a}_{2},\acute{a}_{1}\right)=a_{0}\acute{a}_{0}+a_{0}\acute{a}_{0}+a_{0}\acute{a}_{2}+a_{2}\acute{a}_{0}+a_{2}\acute{a}_{2}+a_{1}\acute{a}_{1}=0$.
\end{center}

Thus $\Psi \left( C^{\bot }\right) \subseteq \Psi \left( C\right) ^{\bot }$. From Lemma \ref{lem:1}, we can verify that $\left\vert \Psi \left( C\right) ^{\bot}\right\vert =\left\vert \Psi \left( C^{\bot }\right) \right\vert$,  which implies that $\Psi \left( C\right) ^{\bot }=\Psi \left( C^{\bot }\right)$.\\

Clearly, $\Psi \left( C\right)$ is self-orthogonal if $C$ is self-dual. However $\left\vert \Psi \left( C^{\bot }\right) \right\vert =\left\vert\Psi \left( C\right) ^{\bot }\right\vert =q^{3n-k_{1}-k_{2}-k_{3}}$. Hence $\Psi\left( C\right)$ and $\Psi \left( C^{\bot }\right) $ are dual $\F_{q}-$ linear codes.
\end{pf}
\qed
\section{The Weight Enumerators of Linear Codes over $R$}

One of the most important results in coding theory is that MacWilliams identity that describes the connections between a linear code and its dual on the weight enumerator, so we investigate this question over $ \F_{q}+v\F_{q}+v^{2}\F_{q}$.\\
Let $ C $ be a linear code of length $ n $ over $ R $. Supposes that $ a $ is an element of $ R $. For all $ c=\left( c_{0},c_{1},...,c_{n-1}\right)\in R^{n}$, define the weight of $ c $ at $ a $ to be $ w_{a}=\vert\lbrace i\backslash c_{i} =a\rbrace\vert$.\\
\begin{defi}
Let $E_{i}$ be the number of codewords of Lee weight $ i $ in $ C $. Then $ \lbrace E_{0},E_{1},...,E_{3n}\rbrace $ is called the Lee distribution of $ C $.
\end{defi}
Define the lee weight enumerator of $ C $ as $ Lee_{C}\left( X,Y\right)=\underset{c\in C}{\sum }E_{i}X^{3n-i}Y^{i}$. Clearly, $Lee_{C}(X,Y)=\underset{c\in C}{\sum }X^{3n-w_{L}\left( c\right) }Y^{w_{L}\left( c\right)}$. Furthermore, the complete weight enumerator of $ C $ over $ R $ is usually denoted by
\begin{center}
$cwe_{C}\left( X_{0},X_{1},...,X_{\left( q-1\right)^{3}}\right) =\underset{c\in C}{\sum }\left( X_{0}^{w_{g_{0}}\left( c\right)},X_{1}^{w_{g_{1}}\left( c\right) },...,X_{q^{3}-1}^{w_{q^{3}-1}\left( c\right) }\right)$
\end{center}
We denote
$\eta a _{0}=0,\eta _{1}=a_{0},\eta _{2}=a_{1}v,\eta _{3}=a_{2}v^{2},\eta_{4}=a_{0}+a_{1}v,\eta _{5}=a_{0}+a_{2}v^{2},\eta_{6}=a_{1}v+a_{2}v^{2},\eta _{7}=a_{0}+a_{1}v+a_{2}v^{2},$ where $a_{i}\in \F_{q}, i=0,1,2$.\\
For any codeword $ c $ of $ C $, let

\begin{center}
$\begin{array}{ccc}
\alpha _{0}\left( c\right)& =&w_{\eta _{0}}\left( c\right) \\
\alpha _{1}\left( c\right)& =&w_{\eta _{3}}\left( c\right) +w_{\eta_{4}}\left( c\right) +w_{\eta _{6}}\left( c\right) +w_{\eta _{7}}\left(c\right)\\
\alpha _{2}\left( c\right) &=&w_{\eta _{1}}\left( c\right) +w_{\eta_{4}}\left( c\right) +w_{\eta _{5}}\left( c\right) +w_{\eta _{6}}\left(c\right)\\
\alpha _{3}\left( c\right)&=&w_{\eta _{4}}\left( c\right) +w_{\eta_{5}}\left( c\right) +w_{\eta _{7}}\left( c\right)
\end{array}$

\end{center}

The Lee weight $ w_{L}\left( c\right) $ of $ c $ is defined to be
\begin{center}
$ w_{L}\left( c\right) =\alpha _{1}\left( c\right) +2\alpha _{2}\left(c\right) +3\alpha _{3}\left( c\right)$.
\end{center}
We define
\begin{center}
$\begin{array}{ccc}
  swe_{C}\left( X_{0},X_{1},X_{2},X_{3}\right)&=& cwe_{C}\left( X_{0},X_{1},...,X_{q^{3}-1}\right) \\
&=&\underset{c\in C}{\sum }X_{0}^{\alpha _{0}\left( c\right) }X_{1}^{\alpha_{1}\left( c\right) }X_{2^{\alpha _{2}\left( c\right)} }X_{3}^{\alpha_{3}\left( c\right) }.
\end{array}$
\end{center}
The Hamming weight enumerator of a code $ C $ of length $ n $ is defined to be
\begin{center}
$  Ham_{C}\left( X,\text{ }Y\right) =\underset{c\in C}{\sum }X^{n-w_{H}\left(c\right) }Y^{w_{H}\left( c\right) }$
\end{center}
\begin{thm}
Let $ C $ be linear code over $ \F_{q}+v\F_{q}+v^{2}\F_{q} $, then
\begin{enumerate}
\item $Lee_{C}(X,Y)=cwe_{C}\left( X^{3},\text{ }X^{2}Y,\text{ }XY^{2},Y^{3}\right)$.

\item $Ham(X,Y)=cwe_{C}\left( X,Y,Y,Y\right)$.

\item $Lee_{C}(X,Y)=w_{\Psi \left( C\right) }\left( X,Y\right)$.

\item $Lee_{C^{\bot }}(X,Y)=\frac{1}{\left\vert C\right\vert }Lee_{C}(X+Y, X-Y)$.
\end{enumerate}
\end{thm}
\begin{pf}
\begin{enumerate}
\item From the definition of the symmetrized weight enumerator, we have

$\begin{array}{ccc}
 cwe_{C}\left( X^{3},X^{2}Y,XY^{2},Y^{3}\right) &=&\underset{c\in C}{\sum }X^{3\alpha _{0}\left( c\right) }\left( X^{2}Y\right)^{\alpha _{1}\left( c\right) }\left( XY^{2}\right) ^{\alpha _{2}\left(
c\right) }\left( Y^{3}\right) ^{\alpha _{3}\left( c\right) } \\
&=&\underset{c\in C}{\sum }X^{3\alpha _{0}\left( c\right) }X^{2\alpha_{1}\left( c\right) }Y^{\alpha _{1}\left( c\right) }X^{2\alpha _{2}\left(c\right) }Y^{2\alpha _{2}\left( c\right) }Y^{3\alpha _{3}\left( c\right) } \\
&=&\underset{c\in C}{\sum }X^{3\alpha _{0}\left( c\right) +2\alpha _{1}\left( c\right) +\alpha _{2}\left( c\right) }Y^{\alpha _{1}\left(c\right) +}{}^{2\alpha _{2}\left( c\right) +3\alpha _{3}\left( c\right) } \\
&=&\underset{c\in C}{\sum }X^{3n-w_{L}\left( c\right) }Y^{w_{L}\left(c\right) } \\
&=& Lee_{C}(X,Y).
\end{array}$
\item From the definition of symmetrized weight enumerator, we have
\begin{center}
$ \begin{array}{cccc}
 cwe_{C}\left( X,Y,Y,Y\right) &=&\underset{c\in C}{\sum }X^{\alpha _{0}\left(c\right) }Y^{\alpha _{1}\left( c\right) }Y^{\alpha _{2}\left( c\right)}Y^{\alpha _{3}\left( c\right) } \\
&=&\underset{c\in C}{\sum }X^{\alpha _{0}\left( c\right) }Y^{\alpha
_{1}\left( c\right) +\alpha _{2}\left( c\right) +\alpha _{3}\left( c\right) }\\
&=&\underset{c\in C}{\sum }X^{n-w_{H}\left( c\right) }Y^{w_{H}\left(c\right) } \\
&=&Ham(X,\text{ }Y).
\end{array}$
\end{center}
\item From the definition of Lee weight enumerator, we can obtain
\begin{center}
$ \begin{array}{cccc}
Lee_{C}(X,Y) &=&\underset{\Psi \left( c\right) \in \Psi \left(C\right) }{\sum }X^{3n-w_{L}\left( \Psi \left( c\right) \right)}Y^{w_{L}\left( \Psi \left( c\right) \right) } \\
&=&w_{\Psi \left( C\right) }\left( X,Y\right).
\end{array}$
\end{center}
 \item From theorem 6, we have $ \Psi \left( C\right) ^{\bot }=\Psi \left(C^{\bot }\right)$, and they are linear codes from Lemma 5, we have
 \begin{center}
 $ w_{\Psi \left( C^{\bot }\right) }(X,Y)=\frac{1}{\left\vert \Psi\left( C\right) \right\vert }w_{\Psi \left( C^{\bot }\right) }(X+Y,X-Y)$.
 \end{center}
 On the other hand, since $ \left\vert \Psi \left( C\right) \right\vert
=\left\vert C\right\vert $, and by 3 in this theorem, we have
\begin{center}
$\begin{array}{ccc}
Lee_{C^{\bot }}(X,Y) &=&w_{\Psi \left( C^{\bot }\right) }(X,Y) \\
&=&\frac{1}{\left\vert \Psi \left( C\right) \right\vert }w_{\Psi \left(C^{\bot }\right) }(X+y,X-Y) \\
&=&\frac{1}{\left\vert C\right\vert }Lee_{C}(X+Y,X-Y).
\end{array}$
\end{center}
 \end{enumerate}
\end{pf}
\qed
\section{Cyclic Codes over $R$}

Cyclic codes play a very important role in the coding theory. We give in this section some useful results on cyclic codes over $\F_{q}+v\F_{q}+v^{2}\F_q $ is said to be cyclic if it satisfies:
\begin{center}
$  \left( c_{n-1},c_{0},...,c_{n-2}\right) \in C\text{ whenever }\left(c_{0},c_{1},...,c_{n-1}\right) \in C$.
\end{center}
It is well known that cyclic codes of length $ n $ over $R $ can be identified with an ideal in the quotient ring $R\left[ x\right] \left/ \left\langle x^{n}-1\right\rangle\right.$ via the $ R $-module isomorphism as follows:
 \begin{center}
 $\begin{array}{cccc}

   R^{n} &\rightarrow &R\left[ x\right] \left/ \left\langle
x^{n}-1\right\rangle \right. \\
\left( c_{0},c_{1},...,c_{n-1}\right) &\mapsto
&c_{0}+c_{1}x+...+c_{n-1}x^{n-1}
 \end{array}$
 \end{center}

 \begin{thm}
 A linear code $C=vC_{1}\oplus \left( 1-v\right) C_{2}\oplus \left( 1-v^{2}\right) C_{3}$ is cyclic over $ R $ if and only if  $ C_{1} $, $ C_{2} $, and $C_{3}$ are cyclic codes of length $ n $ over $ \F_q $.
 \end{thm}
 \begin{pf}
 Let $\left( a_{0},a_{1},...,a_{n-1}\right) \in C_{1},\left(b_{0},b_{1},...,b_{n-1}\right) \in C_{2}$ and $\left(d_{0},d_{1},...,d_{n-1}\right) \in C_{3}$. suppose that $c_{i}=va_{i}+\left(1-v\right) b_{i}+\left( 1-v^{2}\right) d_{i}$ for $i=0,1,...,n-1$. Since $C$ is a cyclic code, it follows that $\left(c_{n-1},c_{0},...,c_{n-2}\right)\in C$. Note that $\left( c_{n-1},c_{0},...,c_{n-2}\right) =v\left(
a_{n-1},a_{0},...,a_{n-2}\right) +\left( 1-v\right) \left(b_{n-1},b_{0},...,b_{n-2}\right) +\left( 1-v^{2}\right) \left(d_{n-1},d_{0},...,d_{n-2}\right)$. Hence $\left(a_{n-1},a_{0},...,a_{n-2}\right) \in C_{1},\\
 \left(b_{n-1},b_{0},...,b_{n-2}\right)\in C_{2}$ and $\left(d_{n-1},d_{0},...,d_{n-2}\right) \in C_{3}$,  which implies that $C_{1}$, $C_{2}$, and $C_{3}$ are cyclic codes over $\F_{q}$.
 Conversely, suppose that $C_{1},$ $C_{2}$, and $C_{3}$ are cyclic codes over $\F_{q}$. Let $\left( c_{0},c_{1},...,c_{n-1}\right) \in C$, where $c_{i}=va_{i}+\left( 1-v\right) b_{i}+\left( 1-v^{2}\right) d_{i}$ for $i=0,1,...,n-1$. Then $\left( a_{0},a_{1},...,a_{n-1}\right) \in C_{1},\left( b_{0},b_{1},...,b_{n-1}\right) \in C_{2}$ and $\left(d_{0},d_{1},...,d_{n-1}\right) \in C_{3}$. Note that $\left(c_{n-1},c_{0},...,c_{n-2}\right) =v\left( a_{n-1},a_{0},...,a_{n-2}\right)
+\left( 1-v\right) \left( b_{n-1},b_{0},...,b_{n-2}\right) +\left(1-v^{2}\right) \left( d_{n-1},d_{0},...,d_{n-2}\right)$. Therefore, $C$ is a cyclic code over $R$.
 \end{pf}
 \qed
 \begin{cor}
 Let  $C=vC_{1}\oplus \left( 1-v\right) C_{2}\oplus \left( 1-v^{2}\right) C_{3}$ be cyclic code of length $ n $ over $ \F_{q}+v\F_{q}+v^{2}\F_q $, then its dual code $C^{\perp}$ is also cyclic and moreover we have
  $C^{\perp}=vC^{\perp}_{1}\oplus \left( 1-v\right) C^{\perp}_{2}\oplus \left( 1-v^{2}\right) C^{\perp}_{3}$.

 \end{cor}
 \begin{cor}
 There exists a self-dual cyclic code of length $ n $ over $R$ if and only if $ q $ is power of $ 2 $ and $ n $ is even.
 \end{cor}
 \begin{pf}
 We know that $C=vC_{1}\oplus \left( 1-v\right) C_{2}\oplus \left( 1-v^{2}\right) C_{3}$. From \cite[Theorem 1]{YanJia} we have $C_{1}$, $ C_{2}$ and $ C_{3}$ are self-dual cyclic code over $ \F_{q} $ if and only if $ q $ is power of $ 2 $ and $ n $ is even. Then we have the result of $ C $.

 \end{pf}
  \qed

 \begin{thm}
 Let $C=vC_{1}\oplus \left( 1-v\right) C_{2}\oplus \left( 1-v^{2}\right)C_{3}$ be a cyclic code of length over $\F_{q}+v\F_{q}+v^{2}\F_{q}$, then $C=\left\langle vC_{1},\left( 1-v\right) C_{2},\left(1-v^{2}\right) C_{3}\right\rangle$, where $C_{i}=\left\langle f_{i}\right\rangle, f_{i}\in \F_{q}\left(i=1,2,3\right), f_{i}\left/ \left\langle x^{n}-1\right\rangle \right.$, and $\left\vert C\right\vert =q^{3n-\left( \deg f_{1}+\deg f_{2}+\deg f_{3}\right) }$.
 \end{thm}
  \begin{pf}
  Assuming that $C_{i}=\left\langle f_{i}\right\rangle, f_{i}\in \F_{q}, \left\vert C_{i}\right\vert =q^{n-\deg f_{i}}\left(i=1,2,3\right)$. It is obvious that $C\subseteq \left\langle vf_{1},\left(
1-v\right) f_{2},\left( 1-v^{2}\right) f_{3}\right\rangle$. Now, let $r=vf_{1}r_{1}+\left( 1-v\right) f_{2}r_{2}+\left( 1-v^{2}\right)f_{3}r_{3}\subseteq \left\langle vf_{1},\left( 1-v\right) f_{2},\left(
1-v^{2}\right) f_{3}\right\rangle$, where $r, r_{i}\in \left( \F_{q}+v\F_{q}+v^{2}\F_{q}\right) \left[ x\right] ,$ $i=1,2,3$.
So there exists $a_{0}, a_{1}, a_{2}\in \F_{q}\left[ x\right]$ such that $r_{i}=vf_{1}a_{1}+\left( 1-v\right) f_{2}a_{2}+\left(1-v^{2}\right) f_{3}a_{3}.$ Hence $r=vf_{1}r_{1}+\left( 1-v\right)
f_{2}r_{2}+\left( 1-v^{2}\right) f_{3}r_{3}=vf_{1}a_{1}+\left( 1-v\right)
f_{2}a_{2}+\left( 1-v^{2}\right) f_{3}a_{3}\in C$. That is $\left\langle vf_{1},\left( 1-v\right) f_{2},\left( 1-v^{2}\right) f_{3}\right\rangle\subseteq C$.
   \end{pf}
   \qed
   \section{ Formally Self-Dual Codes over $R$}
  Formally self-dual binary codes are extensively studied codes. This class of codes plays a very significant role in coding theory both from partial a theoretical points of view.\\
  We study in this section the different methods construction of formally self-dual codes over $R$.\\
  A code is called self-dual if $C=C^{\perp }$. It is called isodual if $C$ is equivalent to $C^{\perp }$.
   The code $C$ is called formally self-dual if $w_{C}\left( y\right) =w_{C^{\perp }}\left( y\right)$.\\
   And here we present three kinds of construction methods for formally self-dual over $\F_{q}+v\F_{q}+v^{2}\F_{q}$ Specially we obtain the following results.
   \begin{thm}
(consrtuction A) Let $A$ be an $n\times n$ matrix over $\F_{q}+v\F_{q}+v^{2}\F_{q}$ such $A^{T}=A$. Then the code generated $G=\left[ I_{n}\left\vert A\right. \right] $ is an isodual code and hence a formally self-dual code of length $2n$.
\end{thm}
\begin{pf}
 Let
\begin{center}
$ G=\left[ I_{n} \left\vert  A\right. \right] $\\
\end{center}
\begin{center}
$ G=\left(
\begin{array}{ccccccc}
1 & 0 & ...0 & A_{11} & A_{12} & ... & A_{1n} \\
0 & 1 & ...0 & A_{21} & A_{22} & ... & A_{2n} \\
\vdots & \vdots & \vdots & \vdots & \vdots & \vdots & \vdots \\
0 & 0 & ...1 & A_{n1} & A_{n2} & ... & A_{nn}
\end{array}
\right) $

\end{center}
The matrix generated of type $ 2n\times n $.\\
And consider the matrix
\begin{center}
$ \acute{G}=\left[ -A^{T} \left\vert I_{n}\right.\right]$
  \end{center}
\begin{center}
$\acute{G}=\left(
\begin{array}{cccccccc}
-A_{11} & -A_{21} & ...-A_{n1} & 1 & 0 & ... & 0 & 0 \\
-A_{12} & -A_{22} & ...-A_{n2} & 0 & 1 & ... & 0 & 0 \\
\vdots & \vdots & \vdots & \vdots & \vdots & ... & \vdots & \vdots \\
-A_{1n} & -A_{2n} & ...-A_{nn} & 0 & 0 & ... & 0 & 1\end{array}\right) \text{ of type }2n\times n$.

\end{center}
$G$ and $\acute{G}$ generate codes with free rank $k\times n$. We need to show that $\acute{C}=C^{\bot }$. Let $u$ the first row of $G$ and let $v$ the first row of $\acute{C}$, we have $\left\langle u,v\right\rangle =A_{11}-A_{11}=0$.\\
for $u$ the first row of $G$ and $j-th$ row of $\acute{G}$, we have $\left\langle u,v\right\rangle =A_{1n}-A_{1n}=0$. For $u$ the $i-th$ row of $G$ and $j-th$ row of $\acute{G}$ we have $\left\langle u,v\right\rangle _{k}=A_{ij}-A_{ji}=0$. Since $A^{T}=A $. There for $\acute{C}=C^{\bot }$ and $C$ is equivalent to $C^{\bot }.$ Since $w_{L}(-a)=w_{L}(a) $ for all $a\in \F_{q}+v\F_{q}+v^{2}\F_{q},$ this is a weight preserving equivalence.

\end{pf}
\qed
\begin{ex}
Let $ q=3 $ and $ n=5 $ and $ A $ be the matrix
\begin{center}

$ A=\left(
\begin{array}{ccccccc}
0&v&2+v&1+2v+2v^{2}&2v+2v^{2}\\
v&2v+2v^{2}&2&1+v&1+v^{2}\\
2+v&2&2v^{2}&2+v+v^{2}&1+2v\\
1+2v+v^{2}&1+v&2+v+v^{2}&1&v\\
2v+2v^{2}&1+v^{2}&1+2v&v&2\\
\end{array}
\right)  $\\

we have $ A=A^{\top} $. Then $ \left[ I_{5}\left\vert A\right.\right] $ generates a formally self-dual code of length $ 10 $ over $ \F_{q}+v\F_{q} +v^{2}\F_{q}$ and the Gray image of the code is a $ \left[ 30,15,9\right] _{3} $ formally self-dual.

\end{center}
\end{ex}
\begin{thm}
(construction B)  Let $M$ be a circulant matrix  over $\F_{q}+v\F_{q}+v^{2}\F_{q}$ of order $n$. Then $
G=\left( I_{n}\left\vert M\right. \right) $ generates an isodual code and hence a formally self- dual code over $R$. This is called the double circulant construction.
\end{thm}
\begin{pf}
Consider
\begin{center}

$ M=\left(
\begin{array}{cccc}
 M_{11} & M_{12} & ... & M_{1n} \\
M_{1n} & M_{11} & ... & M_{1n-1} \\
\vdots & \vdots & \vdots & \vdots \\
M_{12} & M_{13} & ... & M_{11}
\end {array}
\right) $

\end{center}
the circulant matrix over $\F_{q}+v\F_{q}+v^{2}\F_{q} $ of $n\times n$.

Let
\begin{center}
$G=\left( I_{n}\left\vert M\right. \right) $
\end{center}
\begin{center}
$ G=\left(
\begin{array}{cccccccc}
1 & 0 & ... & 0 & M_{11} & M_{12} & ... & M_{1n} \\
0 & 1 & ... & 0 & M_{1n} & M_{11} & ... & M_{1n-1} \\
\vdots & \vdots & \vdots & \vdots & \vdots & \vdots & \vdots & \vdots \\
0 & 0 & ... & 1 & M_{12} & M_{13} & ... & M_{11}
\end{array}
\right)$  of type $ 2n\times n $
\end{center}
And let
\begin{center}

$\acute{G} =\left( -M^{T}\left\vert I_{n}\right. \right) $
\end{center}
\begin{center}
$ \acute{G}=\left(
\begin{array}{cccccccc}
-M_{11} & -M_{1n} & ... & -M_{12} & 1 & 0 & ... & 0 \\
-M_{12} & -M_{11} & ... & -M_{13} & 0 & 1 & ... & 0 \\
\vdots & \vdots & ... & \vdots & \vdots & \vdots & ... & \vdots \\
-M_{1n} & -M_{1n-1} & ... & -M_{11} & 0 & 0 & ... & 1%
\end{array}
\right) \text{ of type }2n\times n $
\end{center}
When $C$ code generated by $G$, and $\acute{C}$ generated by $\acute{G}$. It can similarly be shown that
$ \acute{C}=C^{\bot} $ is equivalent over $\F_{q}+v\F_{q}+v^{2}\F_{q}$ to be code $C^{\prime\prime }$. There is a permutation of rows such that after applying it to $C^{\prime \prime }$, the first column of $ \sigma\left( M^{T}\right)$ is the same as the first column of $ M $. Namely:
\begin{center}
$ (M_{11},M_{21},...,M_{n1})=(M_{\sigma (1)1}^{T},M_{\sigma(2)1}^{T},...,M_{\sigma (n)1}^{T})=(M_{1\sigma (1)},M_{1\sigma(2)},...,M_{1\sigma (n)}) $.
\end{center}
Since the matrix $M$ is circulant, every column of $M$ is then equal to a column of $\sigma (M^{T})$. Then apply the necessary column permutation $\tau $ so that $\tau (\sigma (M^{T}))=M$.  We apply another column
permutation $\rho $ so that $\sigma \left( I_{n}\right) =I_{n}$.  Which means that $C$ is equivalent to
$ \acute{C }=C^{\bot}$, hence the codes are isodual.
\end{pf}
\qed

\begin{ex}
Let $ q=5 $ and $ n=5 $ and $ M $ be a circulant matrix  \\

$M=\left(
\begin{array}{cccccccccc}
3v+2v^{2}&4v&3+2v&1+2v+2v^{2}&2v+3v^{2}\\
2v+3v^{2}&3v+2v^{2}&4v&3+2v&1+2v+2v^{2}\\
1+2v+2v^{2}&2v+3v^{2}&3v+2v^{2}&4v&3+2v\\
3+2v&1+2v+2v^{2}&2v+3v^{2}&3v+2v^{2}&4v\\
4v&3+2v&3+2v&1+2v+2v^{2}&2v+3v^{2}\\
\end{array}
\right)$\\

Then $ \left[ I_{5} \left\vert M\right.\right] $ generates a formally self-dual code of length $ 10 $ over $ \F_{q}+v\F_{q} +v^{2}\F_{q}$ and the Gray image of the code is a $ \left[ 30,15,12\right] _{5} $ formally self-dual.
\end{ex}
\begin{thm}
(construction C) Let $M$ be a circulant matrix over $\F_{q}+v\F_{q}+v^{2}\F_{q}$ of order $n-1$. Then the matrix

\begin{center}
$G=\left( I_{n}\left\vert
\begin{array}{cccc}
\alpha & \omega & ... & \omega \\
\omega &  &  &  \\
\vdots &  & M &  \\
\omega &  &  &
\end{array}%
\right. \right)$
\end{center}

Where $\alpha ,\omega \in \F_{q}+v\F_{q}+v^{2}\F_{q}$, is generator matrix of a formally self-dual code over $\F_{q}+v\F_{q}+v^{2}\F_{q}$.

\end{thm}
\begin{pf}
For

\begin{center}
$M=\left(
\begin{array}{cccc}
M_{11} & M_{12} & ... & M_{1n-1} \\
M_{1n-1} & M_{11} & ... & M_{1n-2} \\
\vdots & \vdots & \vdots & \vdots \\
M_{12} & M_{13} & ... & M_{11}
\end{array}
\right)$
\end{center}
we have

\begin{center}
$G=\left(
\begin{array}{cccccccc}
1 & 0 & ... & 0 & \alpha & \omega & ... & \omega \\
0 & 1 & ... & 0 & \omega & M_{11} & ... & M_{1n-1} \\
\vdots & \vdots & \vdots & \vdots & \vdots & \vdots & \vdots & \vdots \\
0 & 0 & ... & 1 & \omega & M_{12} & ... & M_{11}
\end{array}
\right)  $
\end{center}
Where $\alpha, \omega \in  \F_{q}+v\F_{q}+v^{2}\F_{q}$.

And consider $\acute{G}$ be given as
\begin{center}
$ \acute{G}=\left(
\begin{array}{cccccccc}
-\alpha & -\omega & ... & -\omega & 1 & 0 & ... & 0 \\
-\omega & -M_{11} & ... & -M_{12} & 0 & 1 & ... & 0 \\
\vdots & \vdots & ... & \vdots & \vdots & \vdots & ... & \vdots \\
-\omega & -M_{1n-1} & ... & -M_{11} & 0 & 0 & ... & 1
\end{array}
\right)$.
\end{center}
Let $C=\left\langle G\right\rangle $ and $\acute{C}=\langle \acute{G}\rangle $. Both $C$ and $\acute{C}$ are codes of free rank $n$. Let $ v $ be the first row of $ G $, and let $ w $ be the $ j-th $ row of $ \acute{G}$. We have $ \langle u,v\rangle_{k} =w-w=0$. Let $ v $ be the $ i-th $ row of $ G $, and let $ w $ be the $ j-th $ row of $ \acute{G} $. We have $ \langle u,v\rangle_{k} =M_{ij}-M_{ij}=0$. Hence $ C $ and $ \acute{C}=C^{\perp} $.\\
We see that $ C $ and $ \acute{C} $ have the same weight enumerator. Hence $ C $ and $ C^{\perp} $ have the same weight enumerators.

\end{pf}

\begin{ex}
Let $ q=3 $ and $ \alpha=2+v+2v^{2} $ and $ \omega=2+2v $ and $ n=4 $ and $ M $ be a circulant matrix of order $n-1=3$\\

\begin{center}

$M=\left(
\begin{array}{cccccccccc}
2&1+v&2v^{2}\\
2v^{2}&2&1+v\\
1+v&v^{2}&2\\
\end{array}
\right)$\\
\end{center}

then

\begin{center}
$G=\left(
\begin{array}{cccccccc}
1 & 0 &0 & 0 & 2+v+2v& 2+2v& 2+2v & 2+2v \\
0 & 1 &0 & 0 & 2+2v &2 & 1+v & 2v^{2} \\
0 & 0 & 1 &0 &2+2v & 2v^{2} & 2 & 1+v \\
0 & 0 & 0& 1 &2+2v &1+v & 2v^{2}& 2
\end{array}
\right)  $
\end{center}

generates a formally self-dual code of length $ 10 $ over $ \F_{q}+v\F_{q} +v^{2}\F_{q}$ and the Gray image of the code is a $ \left[ 24,12,9\right] _{3} $ formally self-dual.
\end{ex}

\qed
\section{The Gray Image of Formally Self-Dual Codes}
Self-dual codes are by definition formally self-dual automatically. There exists formally self-dual codes which are not self-dual. In this section, we study formally self-dual codes over rings with a Gray map.\\
A code is called even if the weights of all codewords are even, in otherwise the code is called odd.
\begin{thm}
 If $C$ is a formally self-dual codes over $\F_{q}+v\F_{q}+v^{2}\F_{q}$ then the image under the corresponding Gray map is formally self-dual code.
\end{thm}
\begin{pf}
Follow from theorem 6. Since self-dual codes are formally self-dual.
\end{pf}
\qed
\begin{lem}
The direct product of formally self-dual codes over a ring $\F_{q}+v\F_{q}+v^{2}\F_{q}$ is a formally self-dual code.
\end{lem}
\begin{pf}
Let $C_{1}$ and $C_{2}$ be formally self-dual codes it follows that $w_{C_{1}}\left( y\right) =w_{C_{1}^{\perp }}\left( y\right) $ and $w_{C_{2}}\left( y\right) =w_{C_{2}^{\bot }}\left( y\right) $. Then
\begin{center}

$\begin{array}{cccc}
w_{C_{1}\times C_{2}}\left( y\right) &=&w_{C_{1}}\left( y\right)w_{C_{2}}\left( y\right) \\
&=&w_{C_{1}^{\perp }\times C_{2}^{\bot }}\left( y\right) \\
&=&w_{C_{1}^{\perp }}\left( y\right) w_{C_{2}^{\bot }}\left( y\right)
 \end{array}.  $

\end{center}
Notice that $\left( C_{1}\times C_{2}\right) ^{\perp}=C_{1}^{\perp }\times
C_{2}^{\perp }$, we have that $C_{1}\times C_{2}$ is formally self-dual code.
\end{pf}
\qed
\begin{thm}
 Linear odd formally self-dual codes exist over $\F_{q}+v\F_{q}+v^{2}\F_{q}$ for all lengths.
\end{thm}
\begin{pf}
Notice there are linear odd formally self-dual code of length $1$, so by previous theorem taking direct products of this code we exit with a result that there are odd formally self-dual of all length.
\end{pf}
\qed


\begin{thebibliography}{99}
\bibitem{Bayram} A. Bayram, I.Siap, Cyclic and constacyclic codes over a
non-chain ring, J.Algebra Comb. Discrete App. 1(1). 1-12.

\bibitem{Batoul} A. Batoul, K. Guenda, A. Kaya, B. Yildiz, Cyclic isodual and
formally self-dual codes over $\mathbb{F}_{q}+v\mathbb{F}_{q},$ EJPAM. Vol.
8, No. 1, 2015, 64-80.

\bibitem{Derti} A. Dertli. Y. Cengellenmis, S. Eren, On Quantum Codes Obtained
From Cyclic Codes Over $\mathbb{F}_{2}+v\mathbb{F}_{2}+v^{2}\mathbb{F}_{2},$
arXiv: 1407. 1232v1$\left[ \text{cs.IT}\right] $ 19 jun 2014.

\bibitem{Dougherty} S. T.Dougherty, Formally self-dual codes and Gray mapes,
Thirteenth International Workshop on Algebraic and Combinatorial coding
Theory June 15-21, 2012, Pomorie, Bulgaria.

%\bibitem{Gao} J. Gao, Y.Gao, Some Results on Linear over $Z_{4}+vZ_{4},$
%arXiv: 1402. 6771v1$\left[ \text{cs.IT}\right] $ 27 Feb 2014.
%
\bibitem{Jia} M.Jia, P.Sol\'{e} and B. Wu, Cyclic Codes and the Weight Enumerator
of Linear Codes over $\mathbb{F}_{2}+v\mathbb{F}_{2}+v^{2}\mathbb{F}_{2}$,
App. Comput. Math, V. 12, N 2, 2013.

\bibitem{Kaya} A.Kaya, B, Yildiz, I. Siap, Quadratic Residue Codes Over $%
\mathbb{F}_{p}+v\mathbb{F}_{p}$ And Their Gray Images, arXiv: 1305. 4508v1$%
\left[ \text{cs.IT}\right] $ 20 May 2013.

bibitem{karadeniz} S. karadeniz, S. T.Dougherty, Construction formally self-dual
codes over $R_{k}.$ May 21, 2013.

\bibitem{Liao} D. Liao, Y.Tang, A Class Of Constacyclic Codes Over $R+vR$ and
Its Gray Image, Int. J. Communication, Network and System Sciences, 2012, 5,
222-227.

\bibitem{Zhu} S. Zhu, L. Wang, A class of constracyclic codes over $\mathbb{F}_{p}+v\mathbb{F}_{p}$ and its Gray image, Discrete Mathematics 311(2011)
2677-2682.

\bibitem{Liu} Y. Liu, M. Shi, P. Sol\'{e}, Quadratic residue codes over $\mathbb{F}_{p}+v\mathbb{F}_{p}+v^{2}\mathbb{F}_{p},$ MSC(2010): Primary 94B15; Secondary 11A15.
    
\bibitem{YanJia} Y. Jia, S. Ling and C. Xing, On Self-Dual Cyclic Codes over Finite Fields. IEEE transactions on information theory, Vol 57, O 4, April 2011.
\end{thebibliography}
\end{document}